\documentclass[fleqn,usenatbib]{mnras}

\usepackage{newtxtext,newtxmath}

\usepackage[T1]{fontenc}
\usepackage{ae,aecompl}

\usepackage{graphicx}	
\usepackage{amsmath}	
\usepackage{subfig}
\usepackage{booktabs}
\usepackage{afterpage}
\usepackage{float}
\usepackage{tablefootnote}
\usepackage{threeparttable}

\newcommand{\angstrom}{\mbox{\normalfont\AA}}

\newcommand{\nickel}{$^{56}$Ni }
\newcommand{\mni}{M_\mathrm{Ni56} }

\title[Ca-rich $\gamma$-ray deposition]{The $\gamma$-ray deposition histories of Calcium-rich supernovae}

\author[Sharon \& Kushnir]{
	Amir Sharon$^{1}$\thanks{E-mail: amir.sharon@weizmann.ac.il}
	and Doron Kushnir$^{1}$
	\\
	$^{1}$Dept.of Particle Phys. \& Astrophys., Weizmann Institute of Science, Rehovot 76100, Israel\\
}

\date{Accepted XXX. Received YYY; in original form ZZZ}

\pubyear{2021}

\begin{document}
	\label{firstpage}
	\pagerange{\pageref{firstpage}--\pageref{lastpage}}
	\maketitle
	
	\begin{abstract}
	Calcium-rich supernovae (Ca-rich SNe) are faint, rapidly evolving transients whose progenitor system is yet to be determined. We derive the $\gamma$-ray deposition histories of five Ca-rich SNe from the literature in order to place constraints on possible progenitor systems. We find that the $ \gamma $-ray escape time, $ t_0 $, of the Ca-rich SNe  sample is $\approx35\text{--}65 \,\rm{d}$, within the unoccupied region between Type Ia SNe and stripped envelope supernovae (SESNe). The $ t_0\text{--}\mni $ distribution of these SNe, where $\mni$ is the synthesised $^{56}$Ni mass in the explosion, creates a continuum between the Type Ia and SESNe $ t_0\text{--}\mni $ distribution, hinting at a possible connection between all the events. By comparing  our results to models from the literature, we were able to determine that helium shell detonation models and core-collapse models of ultra-stripped stars are unlikely to explain Ca-rich SNe, since the gamma-ray escape time in these models is smaller than the observed values. Models that agree with the observed $ t_0\text{--}\mni $ distribution are explosions of low mass, $M\approx0.75\text{--}0.8\,M_\odot $, white dwarfs  and core-collapse models of stripped stars with an ejecta mass of $M\approx1\text{--}3\,M_{\odot}$. 
	\end{abstract}
	

\section{Introduction}
\label{sec:intro}
\vspace{1mm}

Calcium-rich supernovae (Ca-rich SNe) are faint, rapidly evolving transients with features of [Ca II] emission in their nebular phase spectra \citep[see, e.g.,][]{Perets2010,Kasliwal2012,De2020}. Their light curves are characterised by relatively short rise and decline times and peak magnitudes of $-14$ to $-16.5$, which are fainter than type Ia SNe and most core-collapse (CC) SNe. They are usually detected in old stellar environments in the far outskirts of their host galaxies, suggesting that they originate from old progenitors that have travelled a great distance from their birth site, or that they occur in very faint systems such as globular clusters \citep[although recent works have failed to detect an underlying host system][]{Lyman2014,Lunnan2017}. The exact progenitor systems of Ca-rich SNe remain unknown. 


Several models have been suggested for their origin and explosion mechanism, such as helium-shell detonations on white dwarfs \citep[WDs;][]{Sim2012,Waldman2011}, a tidal disruption of a WD by a neutron star (NS) or an intermediate-mass black hole \citep{Rosswog2008,Metzger2012}, a CC of ultra-stripped stars \citep{Tauris2015,Moriya2017}, and a merger of a WD with another WD or a NS \citep{Zenati2019,Zenati2020,Pakmor2021,Jacobson2022}. 

One approach to constrain the progenitor system is to compare photometric and spectroscopic observations to radiation-transfer calculations of different progenitor models. \cite{Dessart2015} performed simulations of WD helium-shell detonations and found photometric and spectroscopic similarities with several Ca-rich SNe, although the simulated light curves evolved faster than observations. \cite{Moriya2017} studied the CC of ultra-stripped stars (with ejecta mass $<0.2\,M_\odot$), and found that the simulated light curves are consistent with several Ca-rich SNe, although the spectra agreed with only part of the objects. \cite{Polin2021} calculated the nebular spectra of double-detonation sub-Chandra Type Ia explosions, and found that the results of the low-mass progenitors are similar to Ca-rich events, with a high ratio of observed [Ca II]/[Fe III], despite the small amount of synthesised Ca in the simulation. However, the photospheric phase spectra did not match well with the masses considered in their work ($M_\text{WD}\geq0.7\,M_\odot$).

Another approach is to analyse the environments and kinematics of Ca-rich SNe. The conclusion of these studies is that Ca-rich SNe originate from WDs \citep{Foley2015,Perets2021}, or that the CC of massive stars cannot be their only explosion channel \citep{Dong2022}. \cite{Shen2019} proposed that the progenitors of Ca-rich SNe are either old, metal-poor stars (with an unknown explosion mechanism) or binary systems with at least one white dwarf. Binary systems dynamically form in a globular cluster and are then ejected from the cluster and explode - either due to a helium-shell detonation caused by a merger (for a double-WD binary), helium shell deflagration (for a WD+He-burning star companion), or a tidal disruption (for a WD+NS binary).

The volumetric rate of Ca-rich SNe may also be used to constrain their progenitor system. While the sample of Ca-rich events is quite small, it has increased significantly in the last decade due to large-scale surveys. Using a sample of three events from the Palomar Transient Factory, \cite{Frohmaier2018} calculated a rate of $  1.21_{-0.39}^{+1.13} \times10^4 \,\text{Gpc}^{-3}\,\text{yr}^{-1}  $. The ZTF CLU campaign \citep{De2020} found eight new sources with peak $ r $ band magnitudes $ -17<M_r<-15.5 $, and estimated their volumetric rate to be much smaller, $ 2.21^{+1.01}_{-0.67}\times10^3\,\text{Gpc}^{-3}\text{yr}^{-1} $.
\cite{De2020} also distinguished between two sub-classes of Ca-rich SNe based on the spectroscopic similarity at peak light: Type Ib/c SNe (Ca-Ib/c) and SN 1991bg-like Type Ia SNe (Ca-Ia). They further claimed that the peak light spectroscopic properties form a continuum between the two subtypes. 

\begin{figure*}
	\includegraphics[width=\textwidth]{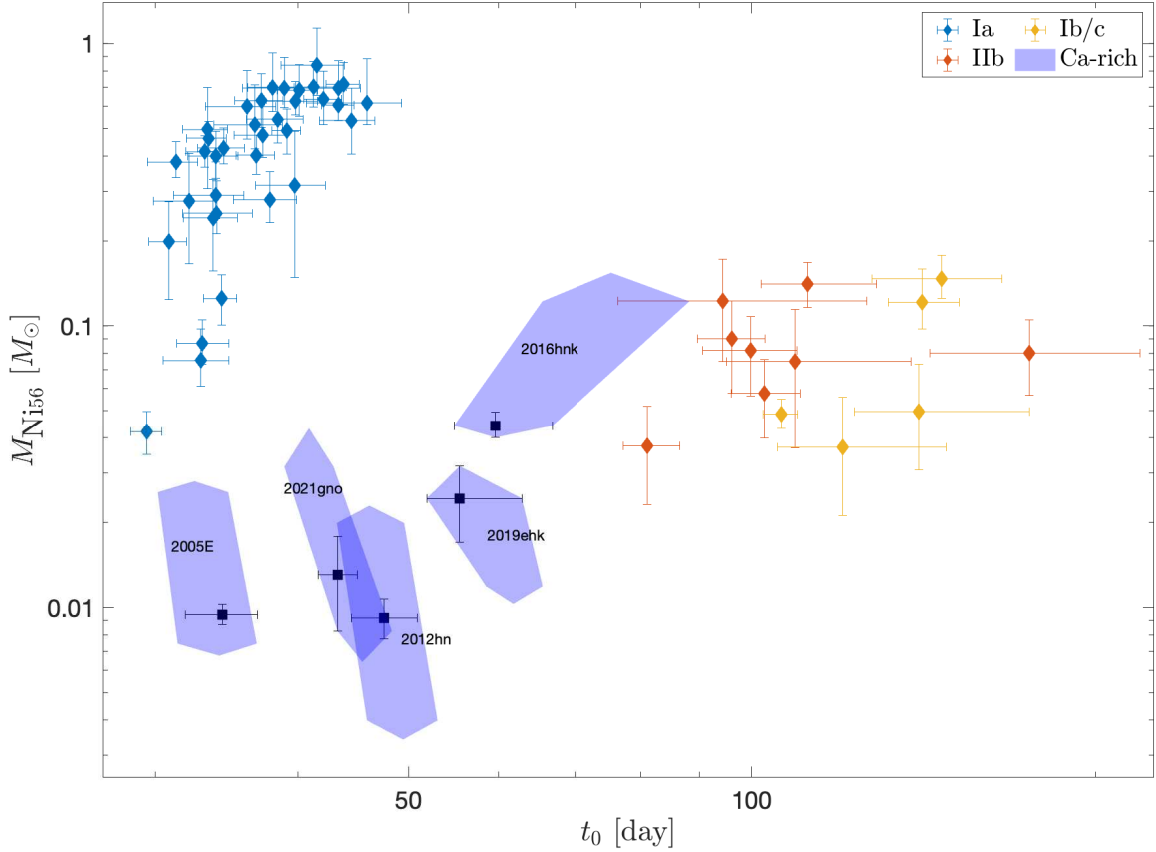}
	\caption{Distribution of the $ \gamma $-ray escape time, $ t_0$, and $\mni$, for the Ca-rich SNe analysed in this work, compared with other types of SNe. Ca-rich SNe (black symbols) have $ t_0 $ values between $35\text{-}60\,\rm{d} $. The blue shaded regions show the $ t_0$-$\mni$ range for different host extinctions ($ 0<E(B-V)<0.5\,\text{mag} $); for SN 2005E, SN 2012hn, and SN 2021gno, a de-reddened NIR correction function was used to correct their bolometric light-curves. The other types of SNe, taken from \protect\cite{Sharon2020}, include Type Ia (blue symbols), Type IIb (orange symbols), and Type Ib/c (yellow symbols). Note that the Ca-rich sample is found in a unique region and does not overlap with the other SN types.}
	\label{fig:t0-ni}
\end{figure*}

In this paper, we adopt an alternative approach. Namely, we constrain the progenitor systems of Ca-rich SNe by studying $ t_0 $, the $ \gamma $-ray escape time of radioactive products through the ejected material, defined by \citep[see][]{Jeffery1999}:
\begin{equation}\label{eq:dep_late}
f_\text{dep}(t) = \frac{t_0^2}{t^2},\;\;\;f_\text{dep}\ll 1,
\end{equation}
where $t$ is the time since explosion and $f_\text{dep}(t)$ is the $\gamma$-ray deposition function, which describes the fraction of the generated $\gamma$-ray energy that is deposited in the ejecta. At late enough times, the ejecta becomes optically thin and the luminosity equals the instantaneous deposition:
\begin{equation}\label{eq:q_total}
Q_\text{dep}(t) = Q_\gamma(t) f_\text{dep}(t)+Q_\text{pos}(t),
\end{equation}
where $Q_\text{dep}(t)$ is the deposited energy in the ejecta from radioactive decay (in this work, we mostly consider $^{56}$Ni and its products), $Q_\gamma(t)$ is the radioactive energy generated from $\gamma$-ray photons and $Q_\text{pos}(t)$ is the kinetic energy of positrons. For a small enough $\gamma$-ray optical depth, each $\gamma$-ray photon has a small chance of colliding with matter, such that the deposition function is proportional to the column density, which scales as $t^{-2}$.

The $\gamma$-ray escape time, together with the synthesized \nickel mass, can be accurately measured from bolometric light curves of SNe, and can be easily calculated for a given ejecta model, without radiation transfer calculations \citep{Wygoda2019}. In \cite{Sharon2020}, it was shown that the observed $t_0-\mni $ distribution can be used to classify different types of SNe. Type Ia SNe have $ t_0\approx30\text{--}45 $, while stripped envelope (SE) SNe have $ t_0\approx80\text{--}170 $.

Here we use the methods of \cite{Sharon2020} to calculate the \nickel mass and the $ \gamma $-ray escape time for Ca-rich SNe. Due to the rarity and dim luminosity of Ca-rich SNe, our sample size is limited and contains only five SNe: SN 2005E, SN 2012hn, SN 2016hnk, SN 2019ehk, and SN 2021gno. The main results are shown in Figure \ref{fig:t0-ni}, where we compare our derived values of $\mni$ and $t_0$ for Ca-rich SNe to other types of SNe. As can be seen in the figure, the Ca-rich sample is located in a region devoid of any other SN types. This unique region forms a continuous bridge between Ia SNe and Type IIb SNe. The $t_0$ values of the Ca-rich sample partially overlap with Ia SNe, with the largest $t_0$ values similar to those of the lowest Type IIb SNe. The $\mni$ values of the Ca-rich sample are, in general, lower than those of other SNe, but the highest $\mni$ values agree with those of Type IIb SNe and with the lowest Ia SNe $\mni$ values. Our findings raise the possibility that Ca-rich SNe are related to either Type IIb SNe or Ia SNe (or both).

The paper is organised as follows: We present the sample of Ca-rich SNe in Section \ref{sec:sample}. In Section \ref{sec:analysis}, we briefly describe the methods of our analysis and the $t_0\text{--}\mni$ distribution of our sample. In Section \ref{sec:models}, we compare our results with models from the literature. We conclude in Section \ref{sec:discussion}.

\section{The C\lowercase{a}-rich SN Sample}
\label{sec:sample}

We analyse five well-observed Ca-rich events - SN 2005E, SN 2012hn, SN 2016hnk, SN 2019ehk, and SN 2021gno. We compute their bolometric light curves using published photometry and estimated reddening and distances, using the methods described in \cite{Sharon2020}. In brief, we first construct light curves in all available bands, where we interpolate and extrapolate for any missing epochs. The sources of photometry are \citet[][SN 2005E]{Perets2010}, \citet[][SN 2012h]{Valenti2014}, \citet[][SN 2016hnk]{Galbany2019,Jacobson2020}, \citet[][SN 2019ehk]{Jacobson2020b,Nakaoka2020}, and \citet[][SN2021gno]{Jacobson2022}\footnote{ \cite{Jacobson2022} also provides the photometric observations of the Ca-rich SN 2021inl, but these observations are not sufficient to construct an accurate light curve with our methods.}. The light curves are converted to flux densities at their effective wavelengths to create a spectral energy distribution (SED) for each epoch. The flux density of wavelengths longer than the band with the longest effective wavelength is estimated with a blackbody (BB) fit, and for short wavelengths it is linearly extrapolated to zero flux at $2000 \angstrom$. Since a significant fraction of the total flux is emitted in near-infrared (NIR) wavelengths (see Section~\ref{sec:IRfrac}), NIR photometry is required to calculate the bolometric light curve accurately. As SN 2005E, SN 2012hn, and SN 2021gno lack NIR observations, we calculate their pseudo-bolometric light-curve ($ 2000<\lambda<8000\,\angstrom $) and correct for the missing NIR flux with the NIR flux fraction of SN 2016hnk, which has the longest time span of NIR observations (see Section~\ref{sec:IRfrac} for a detailed discussion).

It is difficult to estimate the extinction correction of the host galaxy in the case of Ca-rich SNe, a limitation that contributes significantly to the uncertainty of the derived $\mni$ ($t_0$ is less sensitive to the adopted extinction, see Section~\ref{sec:analysis}). We choose a favoured host extinction value for each SNe based on previous works, but we consider for each SNe a wide host reddening range, $ 0<E(B-V)<0.5\,\text{mag} $, with a Milky Way extinction law of $ R_V=3.1 $. Finally, the spectral energy distributions are integrated to obtain the bolometric luminosity. The obtained bolometric light curves, using the favoured extinction values, are shown in Figures \ref{fig:2005E}-\ref{fig:2021gno}. The photometry, the processed photometry (after interpolation, extrapolation, and de-reddening) and the bolometric luminosity of the SNe are included in the supplementary material.

We next describe in more detail each of the SNe in our sample, and in Section~\ref{sec:IRfrac} we discuss the missing NIR flux correcting method.

\subsection*{SN 2005E}
SN 2005E, the Ca-rich SN prototype \citep{Perets2010}, exploded at a projected distance of $ {\sim} 23 $ kpc from the centre of the S0/a galaxy NGC 1032. Spectroscopy at peak light showed similarities with Type Ib SNe. We use the $BVRI$ photometry from \cite{Perets2010}, and we correct for the missing NIR flux using the NIR flux evolution of SN 2016hnk. Following \cite{Perets2010,Waldman2011}, we choose zero host extinction as the favoured value. 

\subsection*{SN 2012hn}
SN 2012hn exploded at a projected distance of 6.2 kpc from the centre of the E/S0-type galaxy NGC 2272. We use photometry from \cite{Valenti2014}, which includes observations from the $ U $ to the $ I $ bands, and we correct for the missing NIR flux using the NIR flux evolution of SN 2016hnk. \cite{Valenti2014} estimated a host extinction of $ E(B-V)_\text{h}=0.2\,\text{mag} $, based on the equivalent width (EW) of Na I D lines, which we adopt as the favoured value.

\subsection*{SN 2016hnk}
SN 2016hnk was located at a projected distance of 3.71 kpc from the centre of the SBa type galaxy MCG-01-06-070 \citep{Galbany2019,Jacobson2020}. The SN showed spectroscopic similarities to SN 1991bg at peak light, but its luminosity decline is slower than Ia SNe and other Ca-rich SNe. 
It was observed in the optical and NIR wavelengths by \cite{Galbany2019} and \cite{Jacobson2020}. On the one hand, \cite{Galbany2019} reported a host extinction of $ E(B-V)_\text{h}=0.45\,\text{mag} $, based on comparisons to Ia SNe and using the observed ratio of H$ \alpha $ and H$ \beta $ fluxes from host-galaxy spectra to estimate the Balmer decrement. \cite{Jacobson2020}, on the other hand, did not correct for host extinction since there was no evidence for Na I D absorption in any spectra. Given that the intrinsic color of Ca-rich SNe is unknown, and given the large uncertainties of the Balmer decrement method of \cite{Galbany2019}, we adopt the value of $ E(B-V)_\text{h}=0\,\text{mag} $ as the favoured value.

\subsection*{SN 2019ehk}
SN 2019ehk exploded close to the core of the star-forming galaxy M100 \citep{Jacobson2020b,Nakaoka2020}. It featured a double-peaked light curve, with the first peak interpreted as the expansion and cooling of a shocked envelope or as a CSM interaction, or a combination thereof \citep{Jacobson2020b}. SN 2019ehk had extensive ground-based optical observations up to $ {\sim}100 $ days after the explosion, which were later supplemented with additional ground-based and HST observations \citep{Jacobson2021,De2021}. NIR observations in the $ JHK $ bands are available up to $ {\sim}25 $ days from the explosion. SN 2019ehk spectra showed a very deep Na I D absorption line, with both \cite{Jacobson2020b} and \cite{Nakaoka2020} estimating a host extinction of $ E(B-V)_\text{h}=0.5\,\text{mag} $, which we adopt as the favoured value. 
\cite{De2021} observed the SN $ {\sim} 280\,$ days from peak light in the $ g $ and $ I $ bands. Using these measurements to calibrate the late-time spectrum in \cite{Jacobson2020}, they measured the [Ca$ \, $II] and [O$ \, $I] line fluxes and interpreted the SN as a core collapse of a $ {\approx}9.5\,M_\odot $ progenitor that was stripped of most of its envelope mass by a companion.

\subsection*{SN 2021gno}
SN 2021gno exploded in the SBa type galaxy NGC 4165, at a projected distance of 3.6 kpc from its centre \citep{Jacobson2022}. Similar to SN 2019ehk, the light curves of SN 2021gno showed two peaks, the first was interpreted to be the result of shock cooling emission and/or a CSM interaction \citep{Jacobson2022}. It was observed from $ {\approx}0.6$ to ${\approx}90 $ days after the explosion with UV and optical filters, but had no NIR observations. We account for the missing flux using the NIR flux of SN 2016hnk. \cite{Jacobson2022} did not correct for host extinction since none of the spectra exhibited Na I D absorption, and we also adopt zero reddening as our favoured value. 

\subsection{NIR fraction}
\label{sec:IRfrac}
\begin{figure}
	\includegraphics[width=\columnwidth]{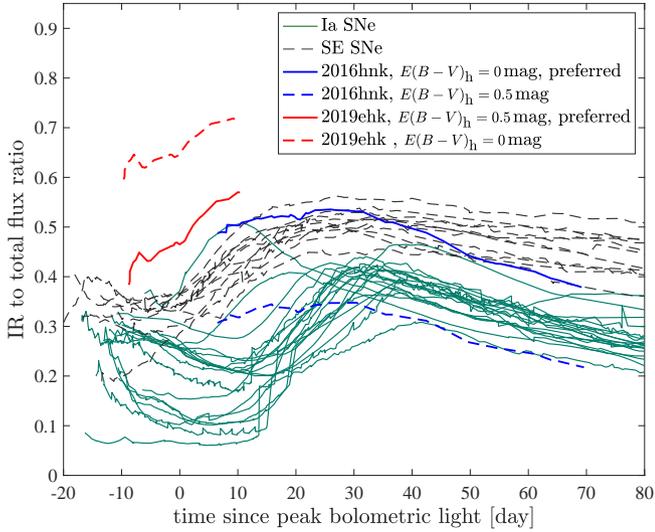}
	\caption{Ratio of the NIR flux ($ \lambda>8000\,\angstrom $) to the total flux as a function of time since peak bolometric light. Blue lines: SN 2016hnk with $ E(B-V)_\text{h}=0\,\text{mag} $  (solid, preferred value) and with $ E(B-V)_\text{h}=0.5\,\text{mag} $ (dashed). Red lines:  SN 2019ehk with $ E(B-V)_\text{h}=0.5\,\text{mag} $  (solid, preferred value) and with $ E(B-V)_\text{h}=0\,\text{mag} $ (dashed). The NIR flux ratios of Ia SNe (solid, teal lines) and SESNe (dashed, black lines) from the sample of \citet{Sharon2020} are shown as well. Assuming that SN 2016hnk and SN 2019ehk share a similar NIR fraction leads to the conclusion that their NIR fraction is similar to SESNe, and that $ E(B-V)_\text{h}\approx0(0.5)\,\text{mag} $ for SN 2016hnk (SN 2019ehk).}
	\label{fig:IRfrac}
\end{figure}

Figure~\ref{fig:IRfrac} shows the ratio of the NIR flux ($ \lambda>8000\,\angstrom $) to the total flux as a function of time since peak bolometric light for SN 2016hnk (solid, blue line) and SN 2019ehk (solid, red line). Since the host extinction estimates are highly uncertain, and in order to demonstrate the effect of the extinction correction, we also plot the NIR fraction of SN 2016hnk with $ E(B-V)_\text{h}=0.5\,\text{mag} $ (dashed, blue line), and of SN 2019ehk with zero host extinction (dashed, red line). The NIR flux ratios of Ia SNe (solid, teal lines) and SESNe (dashed, black lines) from the sample of \cite{Sharon2020} are shown as well. The NIR flux ratio of SN 2016hnk reaches a maximal value of $ {\approx}0.5 $, similarly to SESNe. De-reddening with $ E(B-V)_h=0.5\,\text{mag} $ reduces the maximal value to $ {\approx}0.3 $, in agreement with Type Ia SNe \citep[similar to the $ E(B-V)_\text{h}=0.45\,\text{mag} $ determined by][to match the SN 2016hnk color curve to that of Ia SNe]{Galbany2019}. SNe 2019hnk's NIR ratio, de-reddened  with $  E(B-V)_\text{h}=0.5\,\text{mag} $, is higher than all other SNe, but seems consistent with SN 2016hnk, assuming no host extinction places SN 2016hnk well above the other SNe. Assuming that SN 2016hnk and SN 2019ehk share a similar NIR fraction leads to the conclusion that their NIR fraction is similar to SESNe, and that $ E(B-V)_\text{h}\approx0(0.5)\,\text{mag} $ for SN 2016hnk (SN 2019ehk). This conclusion justifies the adopted NIR correction with the (zero host extinction) NIR flux evolution of SN 2016hnk.


\section{The \lowercase{$ t_0 $}$-M_{\text{N\lowercase{i}}} $ relation of C\lowercase{a}-rich SN\lowercase{e}}
\label{sec:analysis}

In this section, we calculate $t_0$ and $ \mni $ from the bolometric light curves constructed in Section~\ref{sec:sample}. The calculation is based on the Katz integral \citep{Katz2013,Shussman2016,Nakar2016},
described in detail in \cite{Wygoda2019} and \cite{Sharon2020}. In brief, the Katz integral is given by
\begin{equation}\label{eq:integral}
\begin{split}
&QT = LT-ET,\\
&QT \equiv \int_0^t Q_\text{dep}(t')t'dt',\:\:\:&LT \equiv \int_0^t L(t')t'dt',
\end{split}
\end{equation}
where $Q_\text{dep}(t)$ is defined in Equation \eqref{eq:q_total}, $L(t)$ is the bolometric luminosity, and $ET$ is the integrated time-weighted luminosity that would be emitted if no $^{56}$Ni was produced. To describe the deposition fraction at all times, we use the following interpolating function, which connects the expected behaviour at early and late times:
\begin{equation}\label{eq:deposition}
f_\text{dep}(t)=\frac{1}{\left(1+\left(t/t_0\right)^n\right)^{\frac{2}{n}}},
\end{equation}
where $n$ is a parameter that controls the smoothness of the interpolation and is determined in the fitting process.
The quality and time span of the current Ca-rich SN sample are insufficient to accurately determine $ET$, so we assume $ ET=0 $ in what follows. This choice has a negligible impact on $t_0$ and $ \mni $. 
We further omit the first peaks of SN 2019ehk and SN2021gno ($ t<10 $ days since explosion) from our calculations, since it is unlikely that these peaks are $^{56}$Ni-powered \citep{Jacobson2020,Nakaoka2020,Jacobson2022}.

\begin{figure*}
	\includegraphics[width=\textwidth]{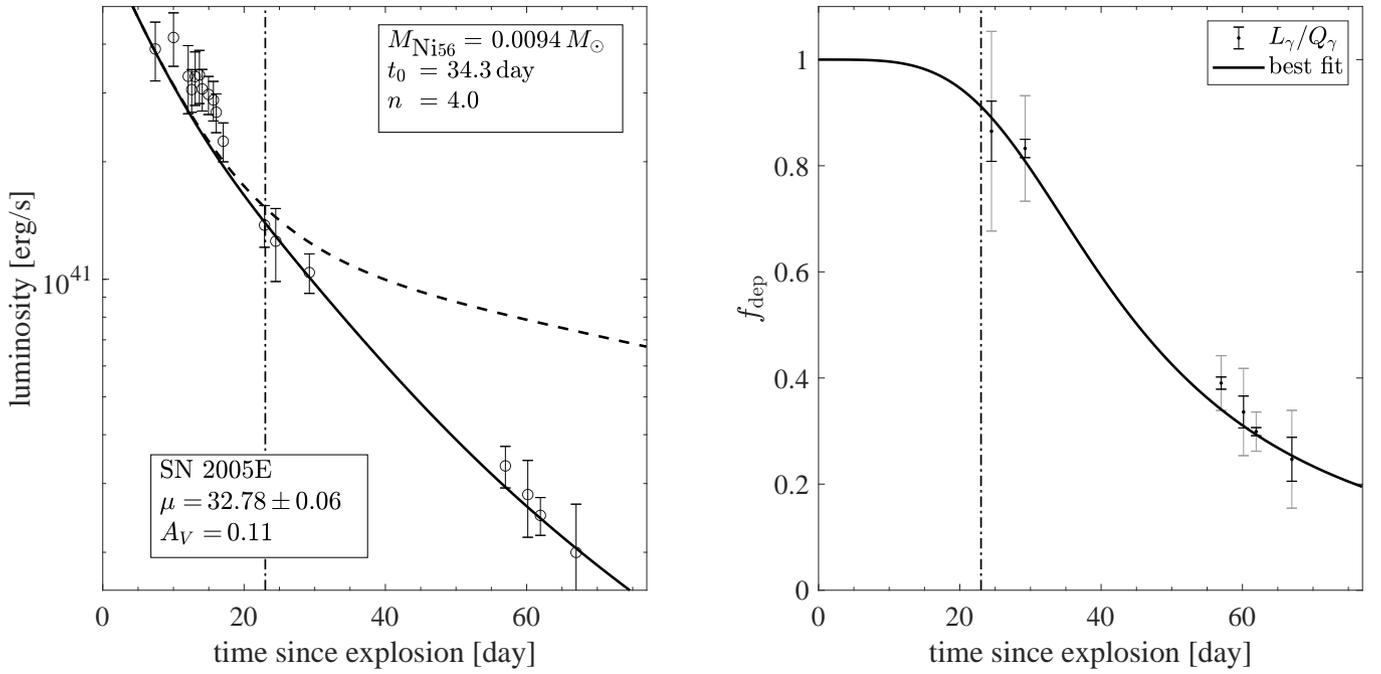}
	\caption{The best-fit results of the Katz integral method for SN 2005E. The median values and 68\% confidence level of the parameters are presented in Table \ref{tab:results}. Left panels: Comparison of the observed bolometric light curves to the best-fit model (with the parameters given in the boxes, solid line) and to the radioactive energy generation rates (same as assuming $f_\text{dep} = 1$, dashed line). The distance and extinction estimates are given in the boxes as well. The errors represent the total errors (statistical and systematic). Right panels: Comparison of the deposition function, $f_{\text{dep}}$, that corresponds to the best-fit model (solid line) to the ratio $(L-Q_{\text{pos}})/Q_{\gamma}$. This ratio corresponds to $L_{\gamma}/Q_{\gamma}$ for $t\in t_{L=Q}$, where we use the observed $L$ and the derived $Q_{\text{pos}},Q_{\gamma}$. The total errors (statistical and systematic) are indicated by grey bars, while the (photometric) statistical errors are indicated by black bars. In both panels, the epochs of $t_\text{min}$ and $t_\text{max}$ (if different from the last phase) are indicated by vertical dashed-dotted lines.}
	\label{fig:2005E}
\end{figure*}

\begin{figure*}
	\includegraphics[width=\textwidth]{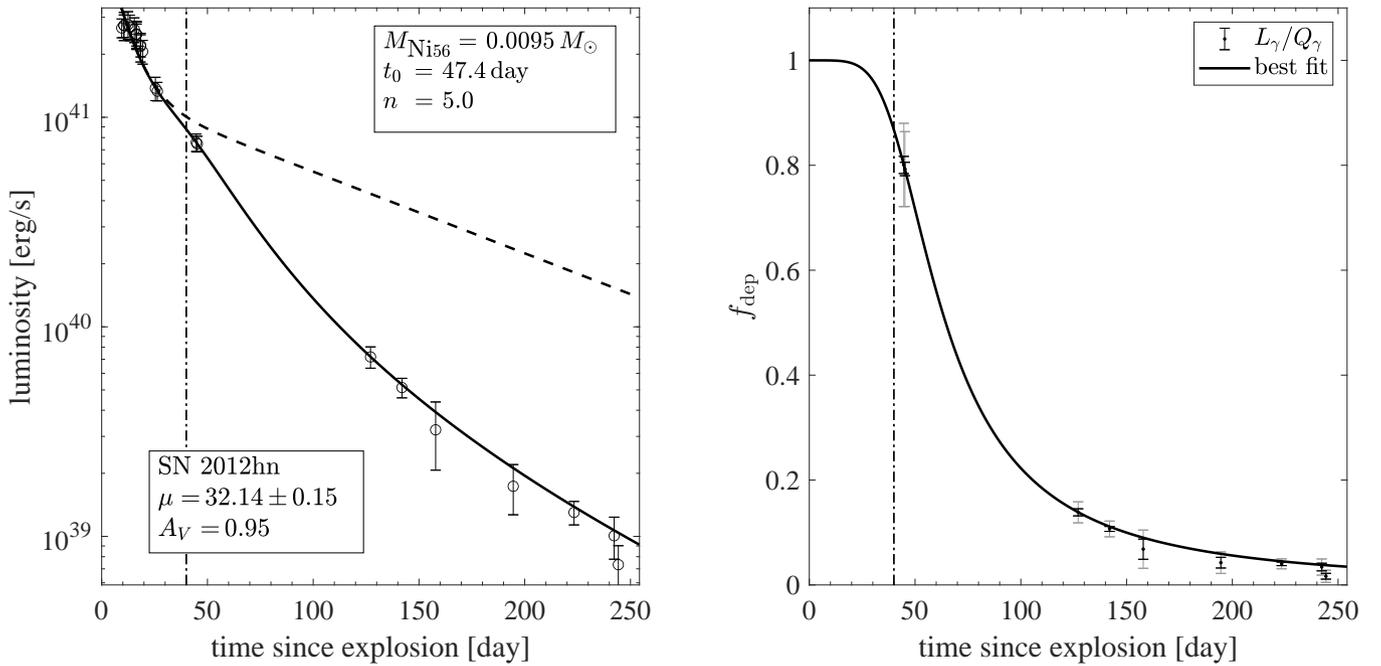}
	\caption{Same as Figure \ref{fig:2005E} for SN 2012hn.}
	\label{fig:2012hn}
\end{figure*}

\begin{figure*}
	\includegraphics[width=\textwidth]{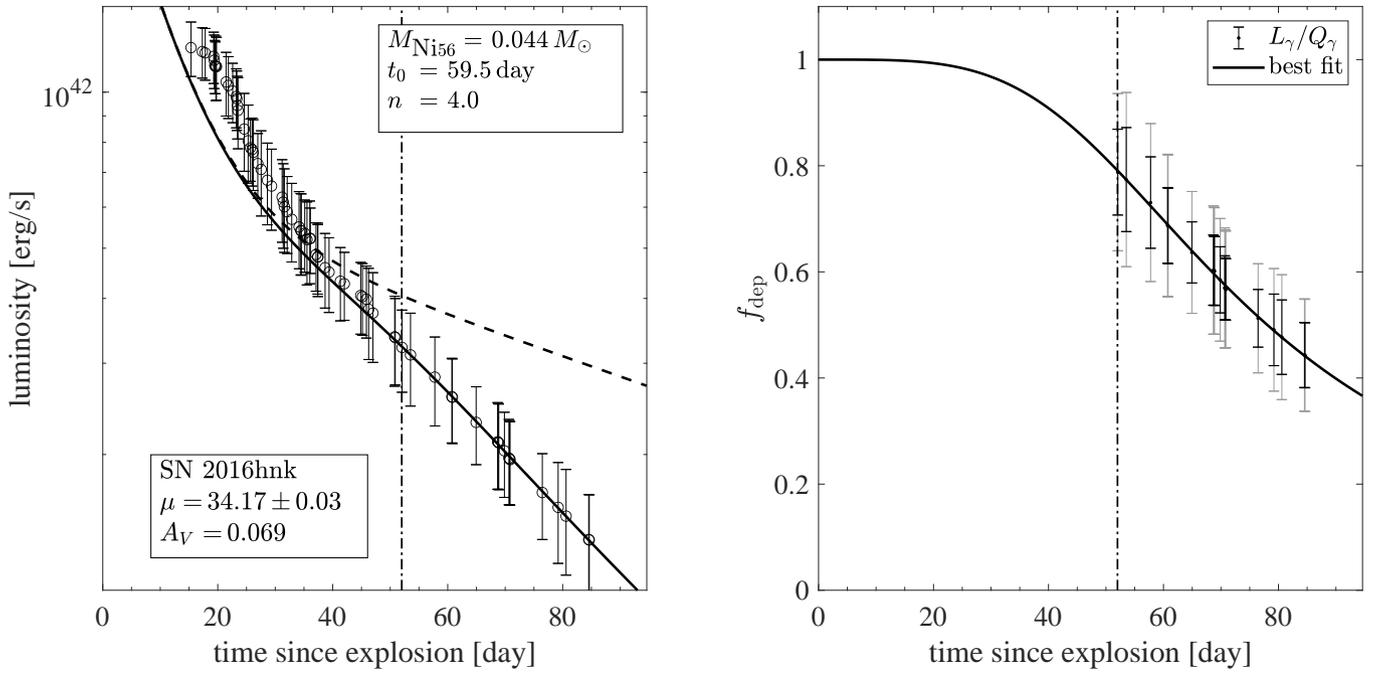}
	\caption{Same as Figure \ref{fig:2005E} for SN 2016hnk.}
	\label{fig:2016hnk}
\end{figure*}

\begin{figure*}
	\includegraphics[width=\textwidth]{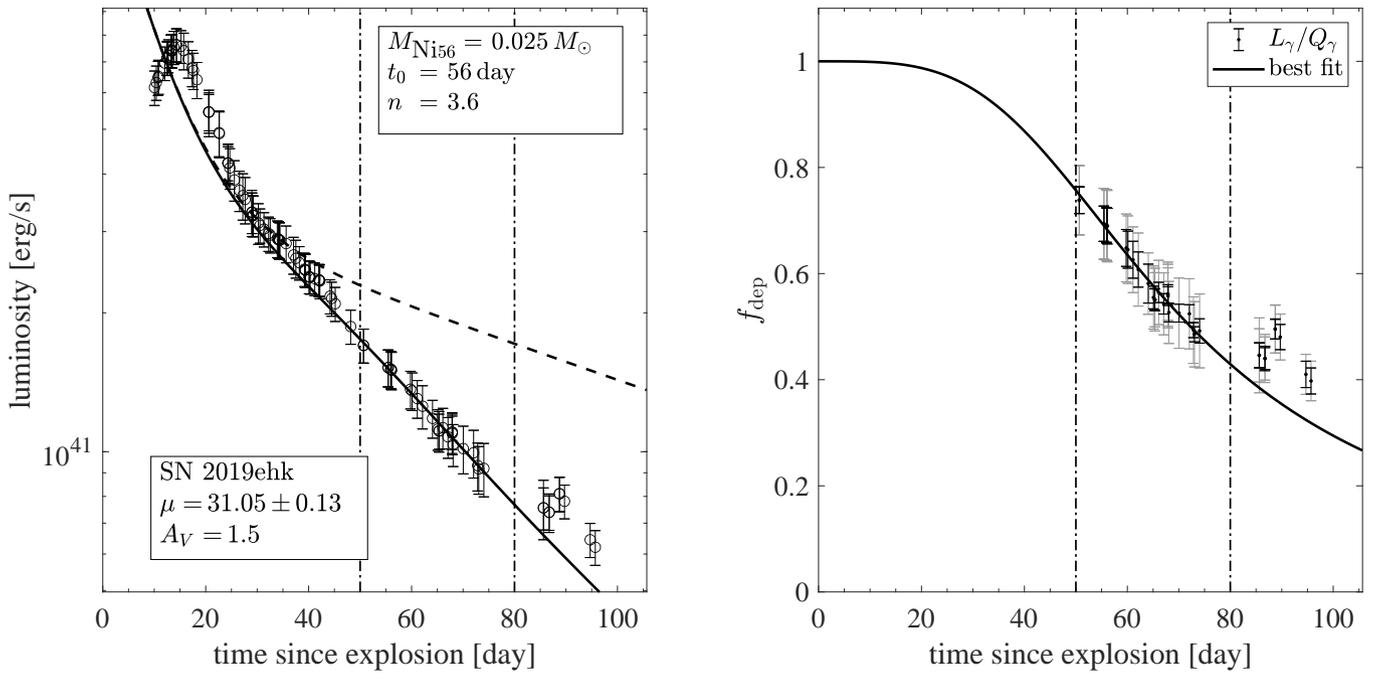}
	\caption{Same as Figure \ref{fig:2005E} for SN 2019ehk.}
	\label{fig:2019ehk}
\end{figure*}

\begin{figure*}
	\includegraphics[width=\textwidth]{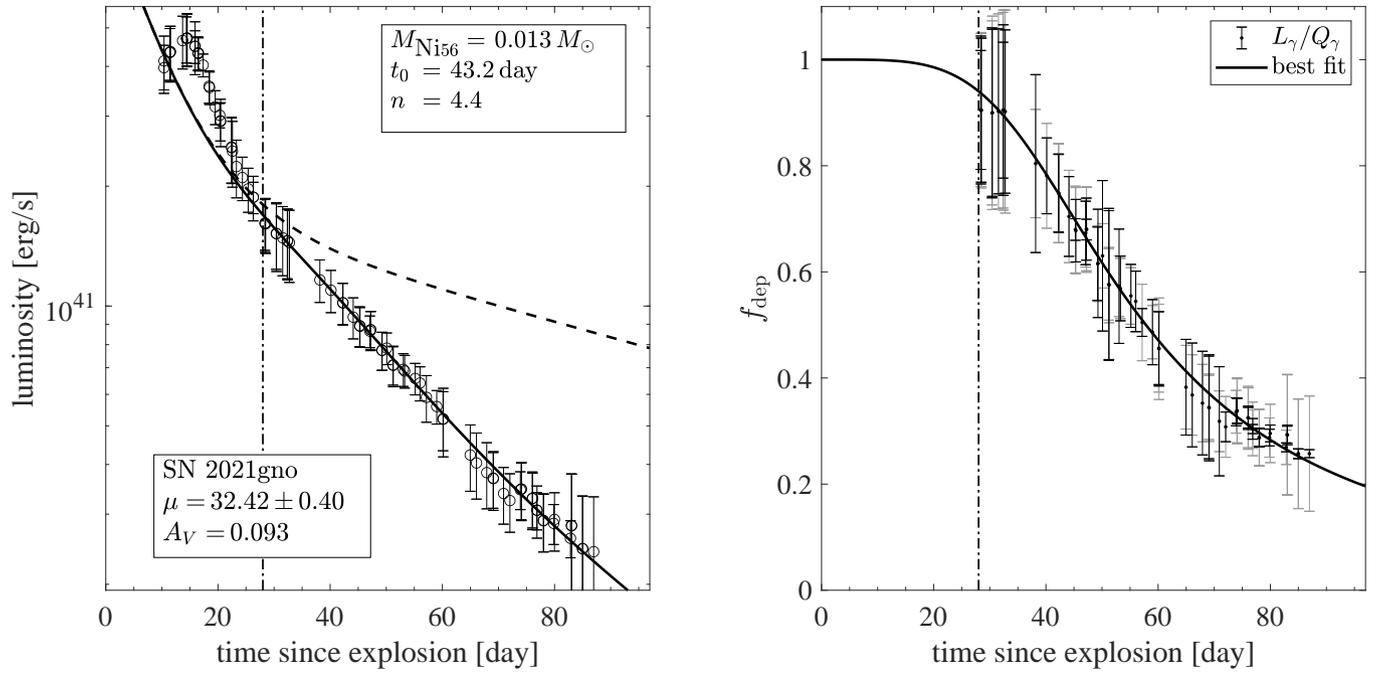}
	\caption{Same as Figure \ref{fig:2005E} for SN 2021gno.}
	\label{fig:2021gno}
\end{figure*}

The fit is performed by minimising the expression 
\begin{equation}\label{eq:likelihood}
\frac{N_\text{bins}}{N_\text{obs}}\sum_{t_i\in t_{L=Q}} \left[\left(\frac{L(t_i)}{LT(t_i)}-\frac{Q_{\text{dep}}(t_i)}{QT(t_i)}\right)\frac{LT(t_i)}{L_{err}(t_i)}\right]^2,
\end{equation}
where $L_\text{err} $ is the luminosity error, $N_\text{obs}$ is the number of observations, and $N_\text{bins}$ is the number of independent time bins, defined as the number of times that $Q_\text{dep}$ changes by $10\%$ over the time range of each SN. The ratio $N_\text{bins}/N_\text{obs}$ affects only the uncertainty of the parameters (and not the best-fit values, see below). The time range $t_{L=Q}=[t_\text{min},t_\text{max}]$ accounts for the times where the assumption of $L=Q_\text{dep}$ is valid. The upper limit, $t_\text{max}$, is determined by the latest epoch where the observations follow the deposition model. In all of the SNe in our sample except SN 2019ehk, $t_\text{max}$ is set to the last phase. SN 2019ehk exhibits a substantial deviation from the deposition model at $t>80\,$day since explosion (see Figure \ref{fig:2019ehk}. The lower limit, $t_\text{min}$, is the earliest epoch at which the fit procedure would result in a deviation of the fit from the observations that is centred around zero. $t_\text{min}$ and $t_\text{max}$ (if different from the last epoch) are indicated in Figures \ref{fig:2005E}-\ref{fig:2021gno} by vertical dashed-dotted lines.
The advantage of using the Katz integral is that Equation \eqref{eq:likelihood} does not depend on $\mni$ and on distance, so the fit is performed over $t_0$ and $n$ alone. $\mni$ is then found by comparing the luminosity in the fitted range to the deposited radioactive energy. 

The uncertainty of the parameters are estimated by performing a Markov Chain Monte Carlo (MCMC) algorithm using the MCMCSTAT Matlab package\footnote{https://mjlaine.github.io/mcmcstat/}, where the likelihood function is Equation~\eqref{eq:likelihood} and the priors are uniformly distributed over reasonable domains.

The inferred $ t_0 $ and $ \mni $ values are given in Table~\ref{tab:results} and shown in Figure~\ref{fig:t0-ni} (black squares). The best-fits to each object are shown in Figures~\ref{fig:2005E}-\ref{fig:2021gno}. We find that the Ca-rich SNe occupy a small region in the $t_0\text{--}\mni$ plane, with $t_0$ values in the range of $35\text{-}65\,\rm{d}$ and $\mni$ values in the range of $(1-5)\times10^{-2}\,M_{\odot}$. 
When comparing our results for SN 2016hnk with the results of \cite{Jacobson2020}, we find that the $t_0$ values are in good agreement, but that our $ \mni $ is $ {\approx}50 $ per cent higher. This difference is expected, since the analysis in \cite{Jacobson2020} is based on a $ 3000\text{--}9000\,\angstrom $ pseudo-bolometric light-curve, and the NIR flux of this SN is ${\approx}40{--}50\%$ of its total flux (see Figure \ref{fig:IRfrac}).

\begin{table*}
	\vspace{-0.25 cm}
	\begin{threeparttable}
		\caption{The bolometric light curve parameters, derived using the Katz integral method. The values of the derived parameters are the median values of the posterior distribution, together with the $68\%$ confidence levels.}
		\renewcommand{\arraystretch}{1.3}
		\begin{tabular}{lccclll}
	Name        &  $ \mu $\tnote{a}  & $ E(B-V)_\text{MW} $\tnote{b}  &$ E(B-V)_\text{host} $\tnote{c}& $ R_V^\text{host} $ & $M_{\text{Ni}56}$ $[M_\odot]$ & $t_0\,$[day]  \\\midrule 
2005E      &  32.78$\,\pm\,$0.06  &  0.03 &  0.00$\,\pm\,$0.00   &  - & $  0.009^{+ 0.001}_{- 0.001}$ & $   34^{+ 3}_{- 3}$\\
2005E, $E(B-V)=0.5$&  32.78$\,\pm\,$0.06  &  0.03 &  0.50$\,\pm\,$0.00   &    2 & $  0.026^{+ 0.002}_{- 0.002}$ & $   32^{+ 2}_{- 2}$\\
2012hn     &  32.14$\,\pm\,$0.15  &  0.10 &  0.20$\,\pm\,$0.05   &  3.1 & $  0.009^{+ 0.001}_{- 0.001}$ & $   48^{+ 3}_{- 3}$\\
2012hn, $E(B-V)=0.5$&  32.14$\,\pm\,$0.15  &  0.10 &  0.50$\,\pm\,$0.00   &  3.1 & $  0.020^{+ 0.003}_{- 0.003}$ & $   46^{+ 3}_{- 3}$\\
2016hnk    &  34.17$\,\pm\,$0.03  &  0.02 &  0.00$\,\pm\,$0.00   &  - & $  0.044^{+ 0.005}_{- 0.004}$ & $   60^{+ 7}_{- 5}$\\
2016hnk, $E(B-V)=0.5$&  34.17$\,\pm\,$0.03  &  0.02 &  0.50$\,\pm\,$0.10   &  3.1 & $  0.122^{+ 0.032}_{- 0.032}$ & $   75^{+13}_{-10}$\\
2019ehk    &  31.05$\,\pm\,$0.13  &  0.02 &  0.47$\,\pm\,$0.10   &  3.1 & $  0.024^{+ 0.007}_{- 0.007}$ & $   55^{+ 7}_{- 4}$\\
2019ehk, $E(B-V)=0$&  31.05$\,\pm\,$0.13  &  0.02 &  0.00$\,\pm\,$0.00   &  - & $  0.012^{+ 0.002}_{- 0.002}$ & $   62^{+ 4}_{- 3}$\\
2021gno    &  32.42$\,\pm\,$0.40  &  0.03 &  0.00$\,\pm\,$0.00   &  - & $  0.013^{+ 0.005}_{- 0.005}$ & $   43^{+ 2}_{- 2}$\\
2021gno, $E(B-V)=0.3$&  32.42$\,\pm\,$0.40  &  0.03 &  0.30$\,\pm\,$0.00   &  3.1 & $  0.032^{+ 0.012}_{- 0.012}$ & $   41^{+ 2}_{- 2}$\\

	\end{tabular}
\begin{tablenotes}
	\item [a] Distance modulus
	\item [b] Galactic extinction towards the SN
	\item [c] Host extinction
\end{tablenotes}
\label{tab:results}
\end{threeparttable}
\end{table*}

Because of the large uncertainty of the host extinction, we repeat the calculations with reddening values in the range of $ 0<E(B-V)<0.5\,\text{mag} $ with a Milky Way extinction law of $ R_V=3.1 $. We also test a different NIR correction for SN 2005E, SN 2012hn, and SN 2021gno, by using the de-reddened NIR fraction of SN 2016hnk. The range of considered host extinctions and NIR corrections change $ \mni $ by up to a factor of $ {\sim}3 $ and $ t_0 $ by up to $ 25 $ per cent from the favoured values (blue-shaded regions in Figure~\ref{fig:t0-ni}). 

We compare in Figure~\ref{fig:t0-ni} the derived $t_0$ and $ \mni $ values of the Ca-rich SNe sample to Ia SNe (blue symbols), Type IIb SNe (red symbols), and Type Ib/c SNe (yellow symbols) samples from \cite{Sharon2020}\footnote{A few additional SNe have been added to the sample: SN 2011fu (IIb), SN 2013aa (Ia), SN 2015bp (Ia), SN 2017cbv (Ia), and SN 2021acat (IIb). Details regarding their analysis will be published in the future.}. The Ca-rich sample is found in a unique region, unoccupied by any other SNe types. This unique region forms a continuous bridge between Ia SNe and Type IIb SNe. The $t_0$ values of the Ca-rich sample partially overlap with those of Ia SNe, with the largest $t_0$ values similar to the lowest Type IIb SNe values. The $\mni$ values of the Ca-rich sample are, in general, lower than the other SNe, but the highest $\mni$ values are in agreement with those of Type IIb SNe and with the lowest Ia SNe $\mni$ values. Our findings raise the possibility that Ca-rich SNe are related to either Type IIb SNe or Ia SNe, or both. We discuss a possible connection between these types of SNe in Section~\ref{sec:discussion}.

%

\section{Comparison with models}
\label{sec:models}

\begin{figure*}
	\includegraphics[width=\textwidth]{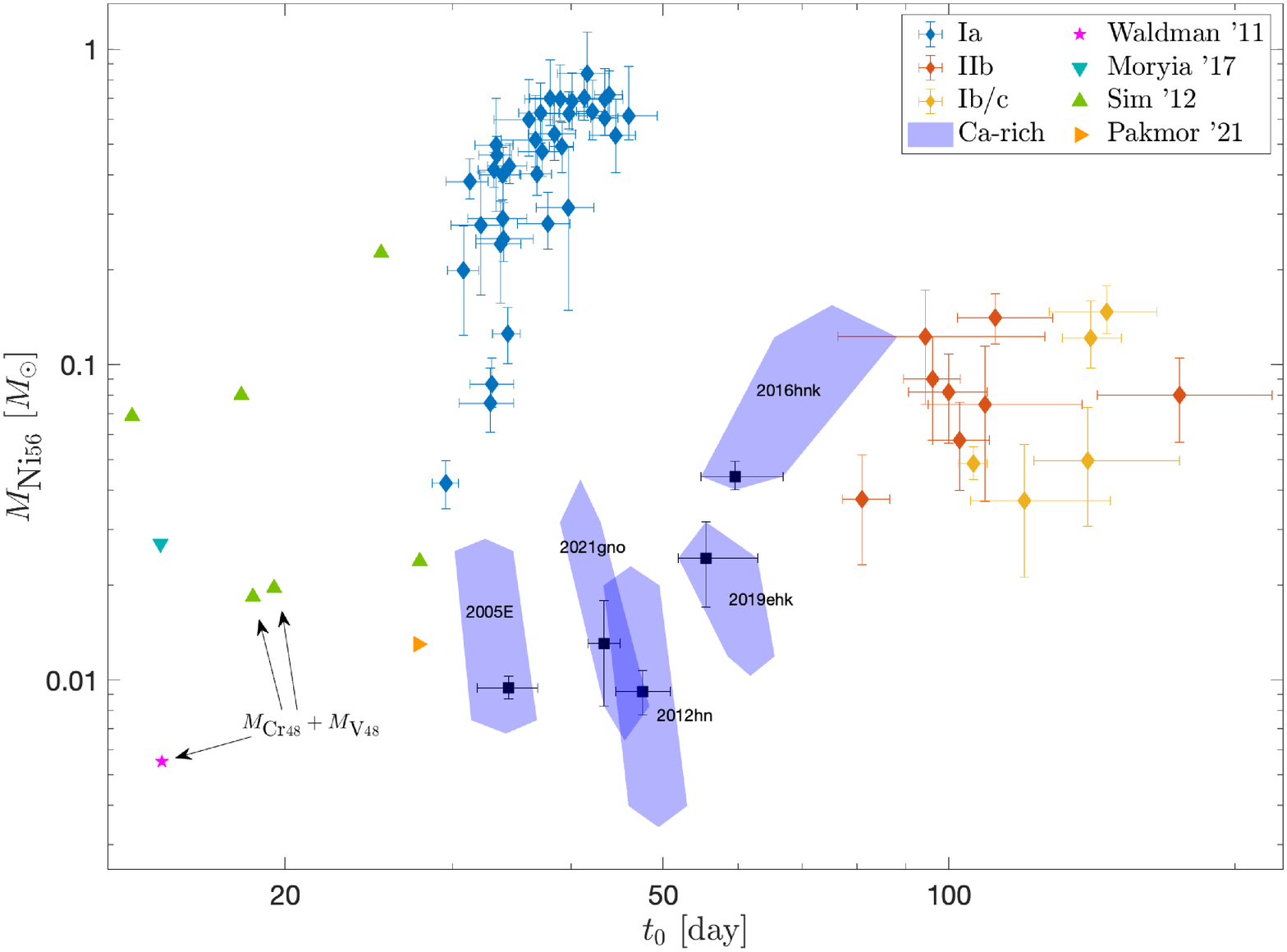}
	\caption{Same as Figure \ref{fig:t0-ni} but superimposed on models of Ca-rich SNe. The models include the helium shell detonations of \protect\cite{Waldman2011} (magenta star), the collapse of ultra-stripped stars of \protect\cite{Moriya2017} (downward-facing, turquoise triangle), the helium shell double-detonations of \protect\cite{Sim2012} (upward-facing, green triangles) and the companion-triggered explosion of a hybrid HeCO WD \protect\cite{Pakmor2021} (rightward-facing, orange triangle). 
 For the model of \protect\cite{Waldman2011} and two of the models of \protect\cite{Sim2012}, $ \mni $ is replaced with $ M_\textrm{Cr48}+M_\textrm{V48} $, since the $ ^{48} $Cr decay chain dominates the energy deposition in these models at $ t\lesssim100 $ days after the explosion.}
	\label{fig:t0-ni_ca}
\end{figure*}

In this section, we compare our observations to models from the literature. We consider both models that were proposed to explain Ca-rich SNe and models of other SNe explosions, specifically Ia SNe and SESNe models. The $\mni$ values of the models are provided in the original works, and we calculate the $t_0$ values of the models by way of $\gamma$-ray MC simulations, using the method in \cite{Sharon2020}\footnote{The explosion ejecta profiles were kindly provided by the authors of the considered publications.}. The results of the Ca-rich models and non-Ca-rich models are presented in Figure \ref{fig:t0-ni_ca} and Figure~\ref{fig:t0-ni_other}, respectively.

We next describe in more detail each of the models.

\subsection{Models of C\lowercase{a}-rich SN\lowercase{e}}
\label{sec:models_Ca-rich}

\subsection*{The helium shell model of \cite{Waldman2011,Dessart2015}}

\cite{Waldman2011,Dessart2015} considered a helium shell detonation at the surface of a low-mass C/O WD \citep[model CO.45HE.2 of][]{Waldman2011}. This model has an ejecta mass of $ 0.2\,M_\odot $, a kinetic energy of $ 0.178\times10^{51}\,\text{erg} $, and its main power source is the decay chain $ ^{48} $Cr$ \rightarrow^{48} $V$ \rightarrow^{48} $Ti (the first step with a half-life of $<1\,\rm{d}$ is not important in what follows). The products of $^{48}$V decay and of $ ^{56} $Co decay have similar $ \gamma $-ray energies, so the $ \gamma $-ray escape opacities are also similar. For a given density profile, we calculate a deviation of  $ {\approx}5 $ per cent in the value of $t_0$ between $^{48}$V and $^{56}$Co. We find for the considered model a small value of $ t_0\approx15\,\text{day} $ (magenta star in Figure~\ref{fig:t0-ni_ca}), as a result of the low ejecta mass. This $t_0$ value is much smaller than the observed $t_0$ values of the Ca-rich SNe sample, so it is likely not a viable explanation for this class of SNe. 

We can also rule out as an explanation for Ca-rich SNe all the explosive shell models that rely on the $ ^{48} $Cr decay chain. Since the half-life of $^{48} $V, ${\approx}15.96\,\rm{d}$, is approximately five times shorter than that of $ ^{56}$Co, the bolometric luminosity of such models drops much faster than all other bolometric light-curves considered in this work, regardless of the $ \gamma $-ray deposition histories. As a result, a fit of the bolometric light curve to $ ^{48} $Cr decay, if such a fit were possible, would lead to a much larger $ t_0 $ than the value obtained for \nickel decay. For example, a full $\gamma$-ray deposition is required at all times (up to $ {\approx}67 $ days after the explosion), for SN 2005E, which results in $ t_0\gtrsim150\,\text{day} $. However, the small ejecta mass of explosive shell models yields a much shorter $ \gamma $-ray escape time. 

\subsection*{The double-detonation of the helium shell of low-mass white dwarfs of \cite{Sim2012}}

\cite{Sim2012} simulated the detonations of an accreted helium layer of $ {\sim}0.21\,M_\odot $ on low-mass carbon-oxygen (CO) WDs. They considered three cases that follow the ignition of the helium shell: a converging-shock double-detonation (CSDD), where the helium detonation compresses the WD core and triggers its detonation near the centre; an edge-lit double-detonation (ELDD), where the CO core is directly ignited at its surface; and a shell-only detonation (HeD), where the CO core fails to ignite. For each type, they considered two initial profiles, 'S' and 'L', with total masses of $ 0.79 $ and $ 0.66\,M_\odot $, respectively. \cite{Sim2012} mention that the helium shell detonation of the L-model is improbable. The simulations' nucleosynthesis yield includes radioactive nuclei other than \nickel, mainly $ ^{48} $Cr and $ ^{52} $Fe, which can overcome the \nickel as the primary power source. This is the case for the ELDD-L and the HeD-L models, where the energy deposition is dominated by $ ^{48} $Cr decay for the relevant times ($ t\lesssim100 $ days since explosion). For each of these models, we calculate the escape time with respect to the $ ^{48} $Cr distribution, although the results are very similar to the escape time with respect to the \nickel distribution.

The $ t_0 $ values of these models are between $ {\sim}14$ to ${\sim}28\,\text{day} $  (upward-facing, green triangles in Figure~\ref{fig:t0-ni_ca}). The highest two values belong to the CSDD models and are close to the lower end of the Ca-rich SNe sample, while the other models have $t_0\lesssim 20\,\text{day} $, which are much smaller than the observations. However, the \nickel mass of the CSDD-S model is higher by several factors than those in all the Ca-rich SNe in the sample, and is an order of magnitude greater than the\nickel mass in SNe with low $ t_0 $. Additionally, the ELDD-L and HeD-L models are primarily powered by $ ^{48} $Cr decay, which further challenges the feasibility of these models. Therefore, out of the six models of \cite{Sim2012}, we find that only the CSDD-L model is somewhat near the observations.
	
\subsection*{The ultra-stripped star model of \cite{Moriya2017}}

We consider the simulated CCSN of an ultra-stripped star with an ejecta mass of $ 0.2\,M_\odot $ and a kinetic energy of $ 0.25\times10^{51}\,\text{erg}  $ from \citep{Moriya2017}. We find a small value of $t_0 \approx15\,\text{day} $ (downward-facing, turquoise triangle in Figure~\ref{fig:t0-ni_ca}), as a result of the low ejecta mass. The $t_0$ value of this model is much smaller than the observed $t_0$ values of the Ca-rich SNe sample, so it is ruled out as a viable explanation for this class of SNe. 

	
\subsection*{The thermonuclear explosion of a massive hybrid HeCO WD triggered by a companion of \cite{Pakmor2021}}

\citet{Pakmor2021} performed a 3D simulation to study the interaction of a He-rich hybrid $0.69\,M_{\odot}$ HeCO WD with a more massive $0.8\,M_{\odot}$ CO WD just before they merge. The accretion from the hybrid WD to the CO WD resulted in a helium detonation that caused the full detonation and disruption of the hybrid WD. A total \nickel mass of $0.018\,M_\odot$ is synthesised in the explosion, out of which $0.013\,M_\odot$ is in the $ 0.6\,M_\odot $ ejecta. The profile we obtained depicts the entire $0.69\,M_{\odot}$ detonated WD, although some of the ejecta would be caught on the unbound, primary WD. We do not attempt to account for this effect, so our results may differ by ${\sim}10\%$ from the actual value. Additionally, the \nickel mass of the profile is lower than that in \cite{Pakmor2021}, ${\approx}0.009\,M_\odot$, which is probably due to remapping between different simulations\footnote{R. Pakmor, private communication.}.
The $t_0$ value of this model is ${\approx} 28\,\textrm{day}$ (right-facing, orange triangle in Figure \ref{fig:t0-ni_ca}), which is higher than that of most WD shell explosion models, but is still somewhat lower than the observations. The synthesised \nickel is in agreement with the fainter Ca-rich SNe.

\subsection{Non-Ca-rich models}

\begin{figure*}
	\includegraphics[width=\textwidth]{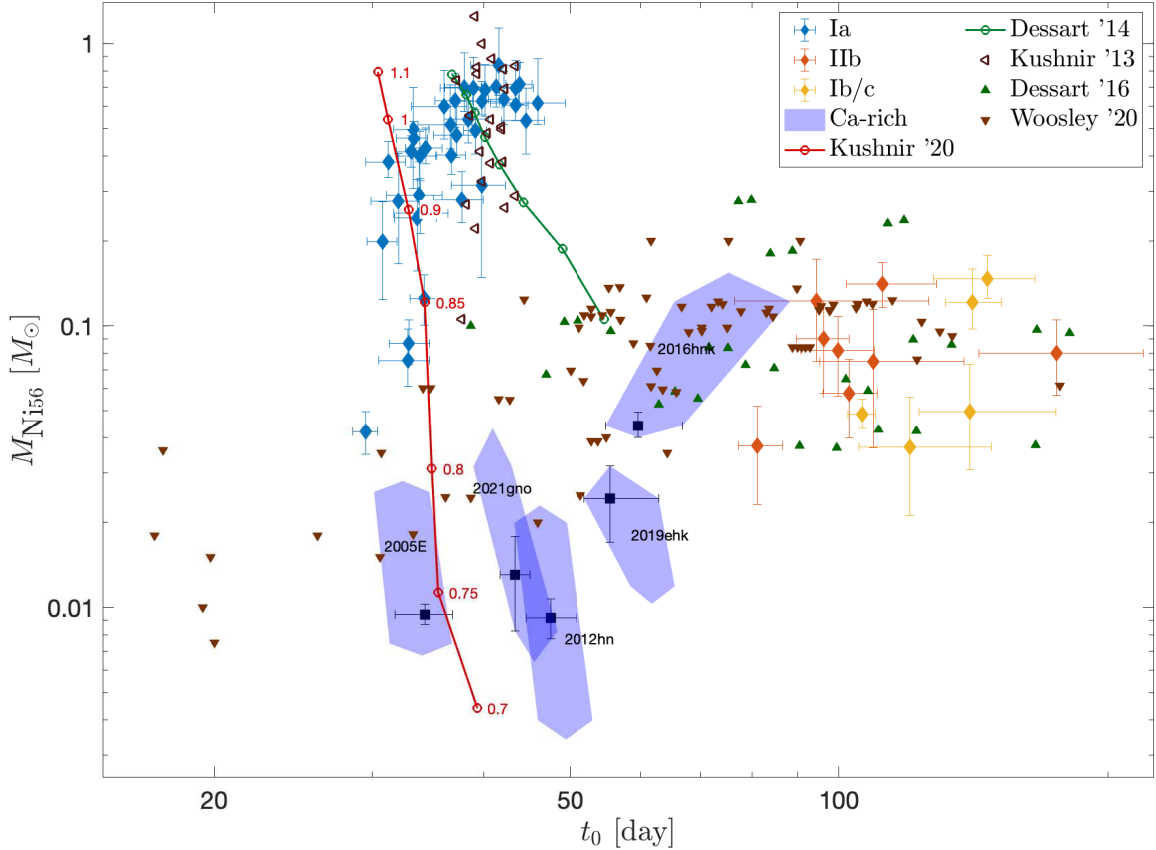}
	\caption{Same as Figure \ref{fig:t0-ni} but superimposed on models of non-Ca-rich SNe. The models consist of Type Ia SNe models; The sub-Chandra detonation of \protect\cite{Kushnir2020} with a progenitor mass range of $0.7\text{--}1.1\,M_\odot$ (red line; progenitor masses are indicated on the plot), Chandrasekhar-mass explosions of \protect\cite{Dessart2015} (green line), and WD collisions of \protect\cite{Kushnir2013} (leftward-facing brown triangles); SESNe: IIb/Ib/Ic explosions of \protect\cite{Dessart2016} (upward-facing, dark-green triangles) and Ib/Ic explosions of stripped helium stars of \protect\cite{Woosley2021} (downward-facing, brown triangles).}
	\label{fig:t0-ni_other}
\end{figure*}

\subsection*{Comparison to Ia SN models}

We consider several Type Ia explosion models, shown in Figure \ref{fig:t0-ni_other}: the sub-Chandra detonation of \cite{Kushnir2020} (red line), Chandrasekhar-mass explosions of \cite{Dessart2015} (green line), and WD collisions of \cite{Kushnir2013} (leftward-facing, brown triangles). The Chandrasekhar-mass explosions and WD collisions do not agree with the Ca-rich observations, but the lower end of the sub-Chandra detonations, with WD masses ${\approx}0.7\text{--}0.8\,M_\odot$, is consistent with the low-$t_0$ Ca-rich SNe. 

\subsection*{Stripped envelope CCSNe models}

Figure \ref{fig:t0-ni_other} also displays stripped-envelope SNe models - IIb/Ib/Ic explosions of \cite{Dessart2016} (upward-facing, dark-green triangles) and Ib/Ic explosions of stripped helium stars of \cite{Woosley2021} (downward-facing brown triangles). Both works span a wide range of ejecta masses and explosion energies. The models of \cite{Dessart2016} have typically higher $t_0$ values than the observations, but there is some overlap of the lower end of the models, corresponding to ejecta masses of ${\approx}1.7\text{--}3.2\,M_\odot$, with the higher end of the observations. The models of \cite{Woosley2021} cover a large region of $\gamma$-ray escape times, including the entire range of the observed Ca-rich $t_0$ values. The profiles that are within the range of the Ca-rich SNe region have helium star masses (ejecta masses) of ${\approx} 2.5\text{--}5.5\,M_\odot$ (${\approx} 0.75\text{--}2.6 \,M_\odot$).

To summarise, we find that the $t_0$ values of most Ca-rich models are too low compared to, and in tension with, the observations, a result of their small ejecta masses. 
We also find that some non-Ca-rich SNe models agree with the observations. Of the models we have considered, we find that low-luminosity thermonuclear WD sub-Chandra detonations \citep{Kushnir2020}, with progenitor masses $ {\lesssim}0.8\,M_\odot$, and stripped envelope CCSNe \citep{Dessart2015,Woosley2021} with ejecta masses of ${\approx}0.5 \text{--} 3\,M_\odot $, have $t_0$ and $\mni$ values that are compatible with the observations.

\section{Discussion}
\label{sec:discussion}

In this work, we calculated the $\gamma$-ray deposition histories of Ca-rich SNe. The $\gamma$-ray escape time, $ t_0 $, and the \nickel mass were previously shown to have different characteristic values for different SN Types \citep{Sharon2020}. The results of this work, shown in Figure \ref{fig:t0-ni}, place Ca-rich SNe in a region unoccupied by other SN types in terms of the $ t_0 $--$ \mni $ distribution. Their $ t_0 $ values fall between those of Ia SNe and SESNe, bridging the gap between the two SN types, though the $ \mni $ values of most of the SNe in the sample are lower than those of Type Ia SNe and SESNe. 

One of the models of Ca-rich SNe is the explosive burning of He shells on WDs, which was proposed by \cite{De2020} to be the strongest candidate for these events due to the early type hosts and high volumetric rates of Ca-rich SNe. By analysing their spectroscopic and photometric properties, they further claimed that Ca-Ia and red Ca-Ib/c events are consistent with a double-detonation of a He shell that ignites the entire star, while green Ca-Ib/c are the results of a shell-only detonation. The red Ca-Ib/c progenitors have a lower total mass and a thicker shell compared to those of Ca-Ia events, and their explosion would result in Fe group elements in the shell and intermediate mass elements in the core. The gamma-ray deposition histories impose several constraints on some of these models. The $ t_0 $ values of the shell-only models of \cite{Waldman2011} and \cite{Sim2012} are inconsistent with observations. In addition, most of them are powered by the $ ^{48} $Cr decay chain, where the half-life of its products is much shorter than the \nickel decay chain. He shell detonations were also studied by \cite{Sim2012}, and we find that only a small part of the models is somewhat close to the observations. We conclude that He shell models are unlikely to explain Ca-rich SNe, since the gamma-ray escape time for these models is smaller than the observed values. Note that unlike optical light-curve comparisons, $\gamma$-ray deposition comparisons rely on known, simple physics.

Explosions of low-mass ($M\approx0.7\text{--}0.8\,M_\odot$) WDs 
are in agreement with the low $t_0$ part of the Ca-rich SNe (see Figure \ref{fig:t0-ni}).
Yet the higher $t_0$ events, i.e., SN 2016hnk and SN 2019ehk, which also have larger $ t_0 $ values, do not agree with these models. 
Some of the properties of SN 2019ehk and SN 2021gno, which are unrelated to our analysis, support a massive star origin for these explosions: SN 2019ehk exploded close to the core of a star-forming galaxy, and both are spectroscopically similar to type Ib and have a double-peaked light curve. In addition, some of the SESN models we have tested are in agreement with the higher $t_0$ events.

Despite the possibility of multiple progenitors, there seems to be a continuity in the  $ t_0 $--$ \mni $ distribution, and the location of the SNe within the distribution is correlated with the continuum of spectroscopic properties, as derived by \cite{De2020}. Three of the SNe in our sample - SN 2005E, SN 2012hn, and SN 2016hnk - were analysed in \cite{De2020}. SN 2005E, showing weak Si II lines, strong He I lines and no line blanketing, is located at one end of the spectroscopic analysis sequence (Ca-Ib/c green objects), and is also at the edge of the $ t_0 $--$ \mni $ distribution, having the lowest $ t_0 $ value. SN 2016hnk, showing strong Si II lines, no He I lines and a line-blanketed continuum, is located at the other end of the sequence of \cite{De2020} (Ca-Ia objects), and is at the other end of the $ t_0 $--$ \mni $ distribution, having the highest $ t_0 $ and $ \mni $ values. SN 2012hn is located in the middle of the sequence (Ca-Ib/c red). The peak spectrum of SN 2019ehk, being similar to that of iPTF12bho \citep{Jacobson2020b}, places the SN in the Ca-Ib/c red region. This is also the case for SN 2021gno, as its spectrum is most similar to that of 2019ehk \citep{Jacobson2022}. All three SNe (i.e., SN 2012hn, SN 2019ehk, and SN2021gno) lie in the middle of the $ t_0 $--$ \mni $ distribution of our analysis. To summarise, the $t_0$ values of our analysis and the position in the classification of \cite{De2020} are highly correlated. However, the SN classified as a Ca-Ia object, SN 2016hnk, is closer to the SESNe distribution than the rest of our sample, while the Ca-Ib/c-classified SNe are farther away from it. 

The continuity in the Ca-rich parameters, which seem to connect the Type Ia and IIb SNe, raises the possibility that the explosion mechanism of these events is similar. Despite the differences between the progenitors of the two types (WD explosions for Type Ia SN vs. the collapse of stripped, massive stars for Type IIb SNe), the energy source of the SN of both types could be the same. While it is well established that the energy source of Type Ia is thermonuclear, the energy source of CCSN is under debate, with the main candidates being gravitational \citep{Woosley2005,Janka2012} or thermonuclear \citep{Burbidge1957,Kushnir2015}. The bridge formed by the Ca-rich SNe that links between Type Ia and IIb SN hints that they might be the result of a similar, though not identical, process.

The newly discovered SN 2022oqm \citep{Irani2022} poses additional challenges for the origin of Ca-rich SNe. Classified as a Ic SNe, its spectrum at ${\sim}60$ days since explosion displays strong C II [and Ca II] emissions with no detectable [O I], marking it as a Ca-rich event. However, \cite{Irani2022} found $t_0$ and $\mni$ values of ${\approx}36$ day and $0.12\,M_\odot$, respectively, placing it within the Type Ia region of the $t_0$--$\mni$ distribution, despite its spectral classification. A massive star origin raises additional difficulties, such as its explosion site properties, though a WD origin faces some challenges as well, and none of the scenarios can be ruled out at present \citep{Irani2022}.

We have also tried to calculate the luminosity function (LF) of Ca-rich SNe using the results from \cite{De2020}, and compared it to the Ia SN LF in \cite{Sharon2021}. The LFs do not overlap, since the Ia LF has a peak $ r $ magnitude $ M_r<-17.5 $, brighter than all of the Ca-rich sample. Comparing the rates of the low end Ia LF with the rates of the high end of the Ca-rich LF might strengthen or rule out the connection between the events. However, we did not find strong evidence in favour of either side, as the rates at both ends have large uncertainties due to the low number of events. Additional events would further constrain the rates at these luminosities.

Uncovering the origins of Ca-rich SNe is a difficult task, as these objects are faint, have lower rates compared to Type Ia and CCSNe, and might be composed of several progenitor systems. The $\gamma$-ray deposition histories analysis might help in this task, and this work has shown that some of the models for the Ca-rich progenitors can indeed be ruled out. However, our analysis is limited due to its small sample size, and might be biased towards luminous events. Additional objects might help reveal the entire range of $ t_0 $ and $ \mni $ of these SNe, allowing to place further constraints on their origin and explosion mechanism.

\section*{Acknowledgements}
 We thank Boaz Katz for useful discussions. DK is supported by a research grant from The Abramson Family Center for Young Scientists, an ISF grant, and by Minerva Stiftung. This work made use of the Heidelberg Supernova Model Archive (HESMA)\footnote{https://hesma.h-its.org}. We thank Luc Dessart, Takashi Moriya, Stuart Sim and Stan Woosley for sharing their ejecta profiles with us.

\bibliographystyle{mnras}
\bibliography{bibliography} 

\bsp	
\label{lastpage}
\end{document}